\documentclass[prl,twocolumn,preprintnumbers,superscriptaddress]{revtex4-2}

\usepackage{graphicx,epstopdf}
\usepackage{amssymb,amsmath}
\usepackage{dcolumn}
\usepackage{bm}
\usepackage{color}
\usepackage{enumerate}
\usepackage[colorlinks=true,citecolor=blue]{hyperref}
\hypersetup{colorlinks=true,citecolor=blue,linkcolor=blue,urlcolor=blue}
\usepackage{soul}

\usepackage{tabularx}
\usepackage[dvipsnames]{xcolor}

\DeclareMathOperator{\sgn}{sgn}

\begin{document}

\title{Exciton localization on a magnetic domain wall in MoSe$_2$-CrI$_3$ heterostructure }

\author{S. Mikkola}
\affiliation{Department of Physics, ITMO University, Saint Petersburg 197101, Russia}

\author{I. Chestnov}
\email{igor\_chestnov@mail.ru}
\affiliation{Department of Physics, ITMO University, Saint Petersburg 197101, Russia}

\author{I. Iorsh}
\affiliation{Department of Physics, ITMO University, Saint Petersburg 197101, Russia}
\affiliation{Abrikosov Center for Theoretical Physics, MIPT, Dolgoprudnyi, Moscow Region 141701, Russia}

\author{V. Shahnazaryan}
\email{vanikshahnazaryan@gmail.com}
\affiliation{Department of Physics, ITMO University, Saint Petersburg 197101, Russia}
\affiliation{Abrikosov Center for Theoretical Physics, MIPT, Dolgoprudnyi, Moscow Region 141701, Russia}

\begin{abstract}
The existence of spontaneous magnetization that fingerprints a ground-state ferromagnetic order was recently observed in two-dimensional (2D) van der Waals materials. 
Despite progress in the fabrication and manipulation of the atom-thick magnets, investigation of nanoscale magnetization properties is still challenging due to the concomitant technical issues. 
We propose a promising approach for a direct visualization of the domain walls formed in 2D magnetic materials.
By interfacing 2D magnet with a transition metal dichalcogenide (TMD) monolayer, the strong proximity effects enable pinning the TMD excitons on the domain wall.
The emergent localization stems from the proximity-induced exchange mixing between spin-dark and spin-bright TMD excitons due to the local in-plane magnetization characteristic of the  domain wall in the magnetic monolayer. 
\end{abstract}

\maketitle

A recent discovery of long-range magnetic order in layered van der Waals (vdW) materials \cite{gong2017,Huang2017} inspires studying emergent magnetic phenomena in low-dimensional systems \cite{Gibertini2019,Burch2018}. 
Due to their exceptional controllability by means of stress, gating or optical excitation, such intrinsic two-dimensional (2D) magnets offer fascinating functionality for the creation of next-generation van der Waals-based spintronic devices \cite{Sierra2021}.

\begin{figure*}
    \centering
    \includegraphics[width=0.95\linewidth]{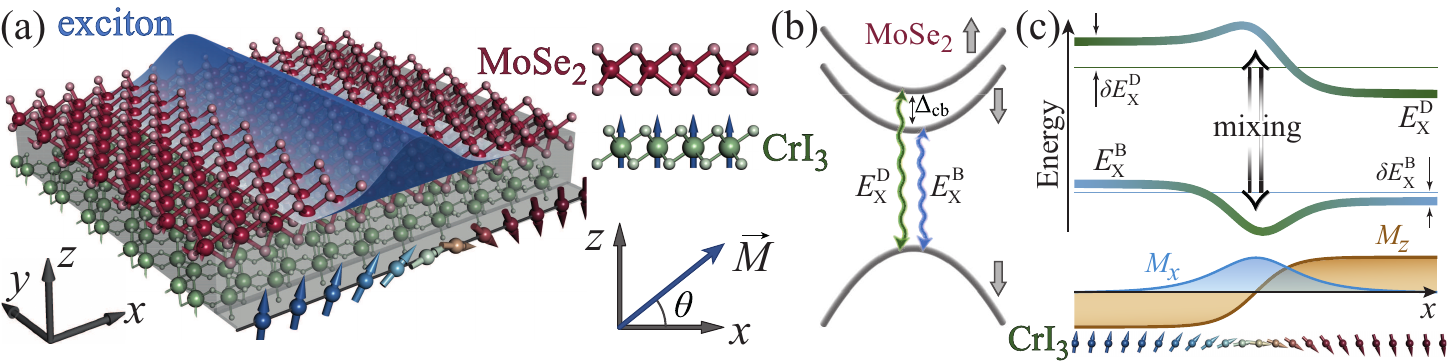}
    \caption{(a) A sketch of the heterostructure. The bottom atom-thick layer of CrI$_3$ contains magnetic domain wall along $y$-direction while the top layer of MoSe$_2$ hosting excitons is covered by a bulk hBN (not shown). 
    The blue dome denotes the exciton state confined over the DW region. The color arrows illustrate spin orientation across DW.
    (b) Schematics of the MoSe$_2$  energy band structure near the $K$-point  which features spin-bright $E_X^B$ and spin-dark $E_X^D$ exciton transitions.
    (c) A sketch of the dark-bright exciton mixing due to in-plane spin orientation in the bottom layer of CrI$_3$. 
    The out-of-plane magnetization $M_z$ (the orange curve) leads to the proximity-induced exchange valley-Zeeman splitting, which pulls exciton energy in opposite directions at the complementary sides of the DW. 
    The in-plane magnetization component $M_x$ locally mixes MoSe$_2$ conduction bands with opposite spins resulting in the exciton level repulsion in the DW core region. The hybridized exciton state is localized at the bottom of the resulting potential trap.
    }
    \label{fig:sketch}
\end{figure*}

Given by intrinsic quantum confinement effects, 2D materials exhibit fascinating properties which emerge at the nanoscale. 
For this reason, manifestations of ferromagnetism in the vdW structures crucially differ from the properties of their bulk counterparts or thin films of conventional ferromagnets. 
One of the most striking differences is the physics standing behind spontaneous formation of a domain structure below Curie temperature. 
The recent theoretical studies \cite{Abdul-Wahab2020,Abdul-Wahab2021} revealed an intricate structure of the interfaces between magnetization domains in vdW magnets. 
In particular, the 2D domain walls (DWs) are exceptionally narrow (down to several nanometers wide) and combine properties of both Bloch and N\'{e}el types.
However, experimental investigation of the nanoscale structure of 2D magnetic DWs remains a challenging problem.
Indeed, a nanoscale spatial resolution is accessible via single-spin magnetometry \cite{Thiel2019,Sun2021,Song2021} realized with nitrogen-vacancy centers in a diamond probe.
Yet, the relevant precision is limited by several tens of nanometers preventing exploration of more fine structures like DWs \cite{Sass2020}.

An alternative approach takes advantage of optically detected proximity effects which arise upon placing 2D magnets in contact with non-magnetic materials. 
For  materials with coupled spin-valley physics such as transition metal dichalcogenide (TMD) monolayers \cite{Xiao2012}, the strong interlayer exchange interaction locks the spin-valley configuration with the direction of magnetization. 
In particular, the out-of-plane magnetization breaks time-reversal symmetry and thus lifts intrinsic valley degeneracy in the TMD monolayer. 
The resulting valley Zeeman splitting is dramatically enhanced by the exchange character of the proximity-induced interactions \cite{Zollner2019,Zollner2023}. 
In CrI$_3$/WSe$_2$ heterostructure the splitting reaches several meV which is equivalent to the unprecedented magnetic field strength up to tens of Tesla \cite{Zhong2017,Seyler2018}. 
Besides, due to the spin-selective interlayer charge transfer \cite{Zhong2017,Zhong2020,Lyons2020}, the TMD photoluminescence exhibits magnetization-dependent degree of circular polarization. 
These effects allow using TMD monolayer as a versatile magnetic sensor to map out microscale magnetization structure of 2D vdW magnets.

However, these techniques do not resolve the effects at the core of DW where magnetization field gradually rotates across the few nanometer-wide interface separating domains. 
In this paper we propose an approach which allows for a direct optical visualization of the DWs in the vdW magnets.  It implies addressing tightly bound TMD excitons whose small spatial extent promotes them as a high-resolution sensor of local magnetization. 
Akin to the real in-plane magnetic field \cite{Molas2017,Zhang2017}, the in-plane magnetization appearing at the DW core mixes spin-polarized TMD sub-bands \cite{Scharf2017} which are naturally split due to the large intrinsic spin-orbit coupling \cite{Wang2018}.
This results in mixing between optically bright and spin-forbidden dark exciton states which correspond to the parallel and antiparallel electron spin configurations, respectively.
The associated local repulsion of exciton energies leads to the exciton confinement at the DW position, which can now be detected by purely optical means.

{\it Model}. -- 
In what follows, we consider excitons in MoSe$_2$ monolayer placed on top of a 2D magnet, see Fig.~\ref{fig:sketch} for the details. As in other Mo-based TMDs, its pristine ground state exciton is spin-bright and optically active, while the spin-dark state lies several meV above, see Fig.~\ref{fig:sketch}b. 
The material parameters and the possible utilization of other TMD monolayers are discussed in the Supplemental Material \cite{SM} (see also Refs. \cite{Rytova1967,keldysh1979,Cudazzo2011,Lau2018,Kormanyos2015,Goryca2019,Yin2021,Echeverry2016} therein).
As a prototypical example of vdW magnetic material we choose a monolayer of chromium triiodide, CrI$_3$, although other chromium trihalides like CrBr$_3$ exhibit rich magnetic domain physics \cite{Lyons2020,Sun2021,Ciorciaro2020} as well.
The temperature of heterostructure is set below the Curie temperature (45 K \cite{Huang2017}) which results in spontaneous formation of the long-range magnetic ordering.
Due to the large magnetic anisotropy of  CrI$_3$, the spins of chromium ions point out-of-the plane \cite{Kim2019} within the regions of homogeneous magnetization.   
In between two antiparallel domains, the unit magnetization vector ${\bm M}$ makes a $\pi$-rotation and hence acquires an in-plane component. 
Without loss of generality we focus on the N\'{e}el-type  DW \cite{Abdul-Wahab2021}  which implies that spins rotate about the $y$-axis parallel to the wall, see Fig.~\ref{fig:sketch}(a).
To be specific, we assume that the $x$-dependence of the spin orientation angle $\theta$ reads
\begin{equation}\label{eq:angle}
    \theta(x) = \arctan(x/l_{\rm dw}),
\end{equation}
where $l_{\rm dw}$ stands for the DW width~\cite{Hubert2008}.

\begin{figure}
    \centering    \includegraphics[width=0.99\linewidth]{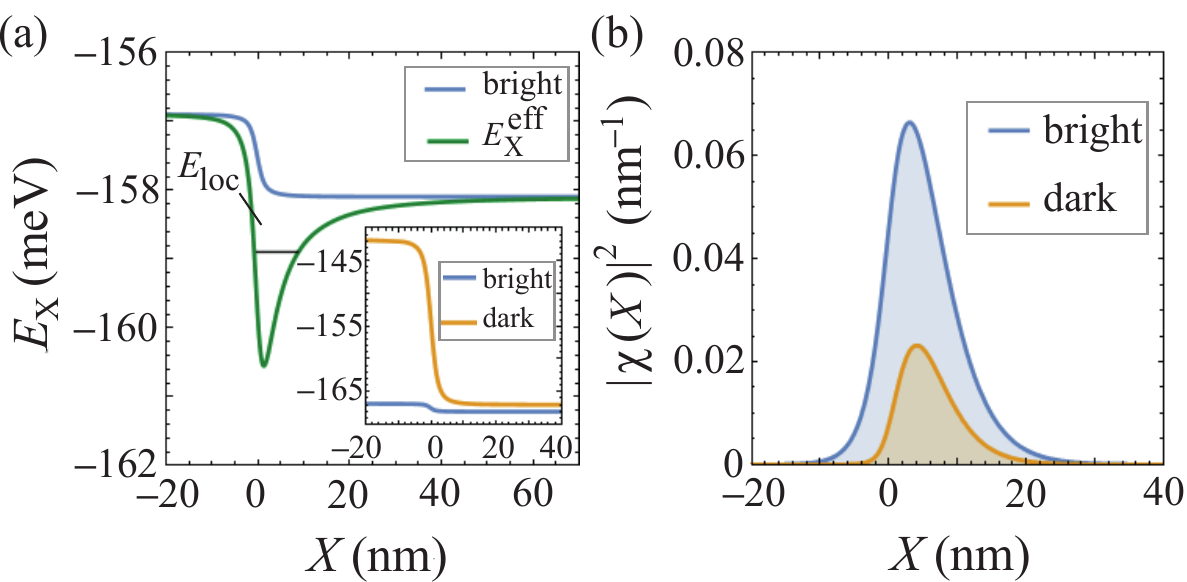}
    \caption{(a) Position-dependent exciton energy $E_{\rm X}^{\rm B}$ (the blue curve), and the effective confinement potential $E^{\rm eff}_{\rm X}$ arising due to the state mixing (the green curve). 
    The energy origin is set to the bottom of MoSe$_2$ conduction band. A thin black line indicates the energy of localized state $E_{\rm loc}$. 
    Inset: Position-dependent energies of the bright (the blue curve), and dark (the orange curve) exciton states affected by the exchange-induced valley-Zeeman splitting. 
    (b) The shape of the hybridized exciton state $\left|\chi_{\rm B[D] }\right|^2$ localized on the magnetic domain wall. Here we set $\alpha_e = 6$~meV, $\alpha_h =6.6$~meV and $l_{\rm dw}=2$~nm. Parameters of the electron-hole relative motion problem which governs $E_{\rm X}^{\rm B}(X)$ and $E_{\rm X}^{\rm B}(X)$ dependencies are given in Ref. \cite{SM}.
    } 
    \label{fig:local}
\end{figure}

Due to proximity exchange effects, both normal $M_z = \sin\left(\theta\right)$ and in-plane $M_x = \cos\left(\theta\right)$  magnetization components affect exciton states in MoSe$_2$. 
In particular, $M_z$ yields a position-dependent energy shift of both spin-bright $E_X^B$ and spin-dark $E_X^D$ exciton species \cite{Zollner2019,Zollner2023}.
However, as ${\bm M}$ acquires an in-plane component, spin is no longer a good quantum number for the conduction and valence band states, resulting in a local mixing between the dark and bright excitons within the same ${K}$ valley \cite{Scharf2017}, similar to the external in-plane magnetic field \cite{Molas2017,Zhang2017}. This effect is sketched in Fig.~\ref{fig:sketch}(c). 
The corresponding effective Hamiltonian in real space reads [see Ref. \cite{SM} for the details]:
\begin{align}
    \label{eq:Ham}
    \hat{H}(X) = 
    \begin{pmatrix}
        -\frac{\hbar^2}{2M_{\rm B}} \frac{{\rm d}^2}{{\rm d} X^2} + E_{\rm X}^{\rm B}(X) & \alpha_e M_x (X) \\ 
        \alpha_e M_x (X) &    -\frac{\hbar^2}{2M_{\rm D}} \frac{{\rm d}^2}{{\rm d} X^2} + E_{\rm X}^{\rm D}(X)
    \end{pmatrix}.
\end{align}
Here $X$, $M_{\rm B [D]}$ are the coordinate and the effective mass of the bright [dark] exciton center-of-mass dynamics. Due to translation invariance along DW, the $Y$-dependence can be factored out.  
$E_{\rm X}^{\rm B[D]}(X)$ are the position-dependent resonance energies of the bright [dark] excitons governed by the local modification of the exciton internal structure in the exchange magnetization field.
The calculated profiles $E_{\rm X}^{\rm B[D]}(X)$ (see Ref. \cite{SM}) which account for the electron-hole binding energies and the proximity-induced valley Zeeman splitting are shown in the inset of Fig.~\ref{fig:local}(a). 
In the homogeneous magnetization field $\left|M_z\right|=1$  away from the DW, the Zeeman energy shift equals $\delta E_{\rm X}^{\rm B} (X)= \sgn(X)(\alpha_e - \alpha_h)$ for the bright exciton and $\delta E_{\rm X}^{\rm D} (X) = -\sgn(X)(\alpha_e  + \alpha_h)$ for the dark state. 
Here $\alpha_{e[h]}$ stands for the spin-independent MoSe$_2$/CrI$_3$ electron [hole] exchange parameter \cite{Zollner2019,Zollner2023}.
Our analysis of the exciton eigenvalue problem yields that at $\alpha_{e,h} < 10$~meV the $E_{\rm X}^{\rm B[D]}(X)$-dependencies can be safely considered as it follows the shape of $M_z(X)$ at the core of the DW [see Ref. \cite{SM} for the details].

The strength of the in-plane magnetization-induced mixing described by the off-diagonal terms in Eq. \eqref{eq:Ham} is governed by the conduction band electron exchange parameter $\alpha_e$ in agreement with Ref.~\cite{Scharf2017}.  
Note that we neglect the back action of the TMD exciton state on the CrI$_3$ monolayer magnetization \cite{Kudlis2021,Dabrowski2022} as well as the state mixing for the TMD valence bands, given by the respective large spin-orbit splitting \cite{Wang2018}. 
The possible impact of exciton states in CrI$_3$ monolayer \cite{Wu2019,Olsen2021,Acharya2022,Kazemi2022} is disregarded as their excitation is suppressed due to the dominant oscillator strength of MoSe$_2$ exciton in the relevant spectral range.

{\it State mixing and exciton localization on the domain wall}. -- 
The exchange-induced mixing at the core of the DW enables exciton localization.
The energy and spatial profile for the ground state $\bm{\chi} = \left( \chi_{\rm B}(X), \chi_{\rm D}(X)  \right)^\intercal$ of Hamiltonian \eqref{eq:Ham} are shown in Fig.~\ref{fig:local}.
The localized ground state has a fraction of dark exciton as a direct implication of local mixing, see Fig.~\ref{fig:local}(b).
We note that the effective potential acting the lower hybridized exciton state is
\begin{equation} \label{eq:EEff}
    E_{\rm X}^{\rm eff} = \frac{1}{2}\left( E_{\rm X}^{\rm B} + E_{\rm X}^{\rm D} 
    -\sqrt{ (E_{\rm X}^{\rm B} - E_{\rm X}^{\rm D} )^2 + 4 \alpha_e^2 M_x^2 } \right),
\end{equation}
which represents an eigenvalue of the potential energy terms in Eq.~\eqref{eq:Ham}. It features a potential trap at the DW position as shown in Fig.~\ref{fig:local}(a), which qualitatively explains the effect of exciton localization. This trap originates from the local mixing while its depth is governed by the exchange coupling parameters $\alpha_{e,h}$ as well as the conduction band splitting $\Delta_{\rm cb}$, see Fig.~\ref{fig:sketch}(b). 
The mixing is maximized at the DW center, $X=0$, where its efficiency is governed by the ratio $\alpha_e/|E_{\rm X}^{\rm B} - E_{\rm X}^{\rm D}|$.

The emergent asymmetry of the trap shape stems from the different effective $g$-factors of the bright and dark states. 
Given by the antiparallel spin configuration, the dark exciton is more sensitive to the out-of-plane magnetization.
Therefore, on the one side of the DW the dark-bright energy splitting $\left| E_{\rm X}^{\rm B} - E_{\rm X}^{\rm D} \right|$ is reduced thus reinforcing mixing effect, see the inset in Fig.~\ref{fig:local}(a).
On the opposite side exciton states pull apart and quench mixing. 
Note that the shape of the trap is mirror reflected in the ${K}^\prime$ valley \cite{SM}.

\begin{figure}
    \centering   
    \includegraphics[width=0.99\linewidth]{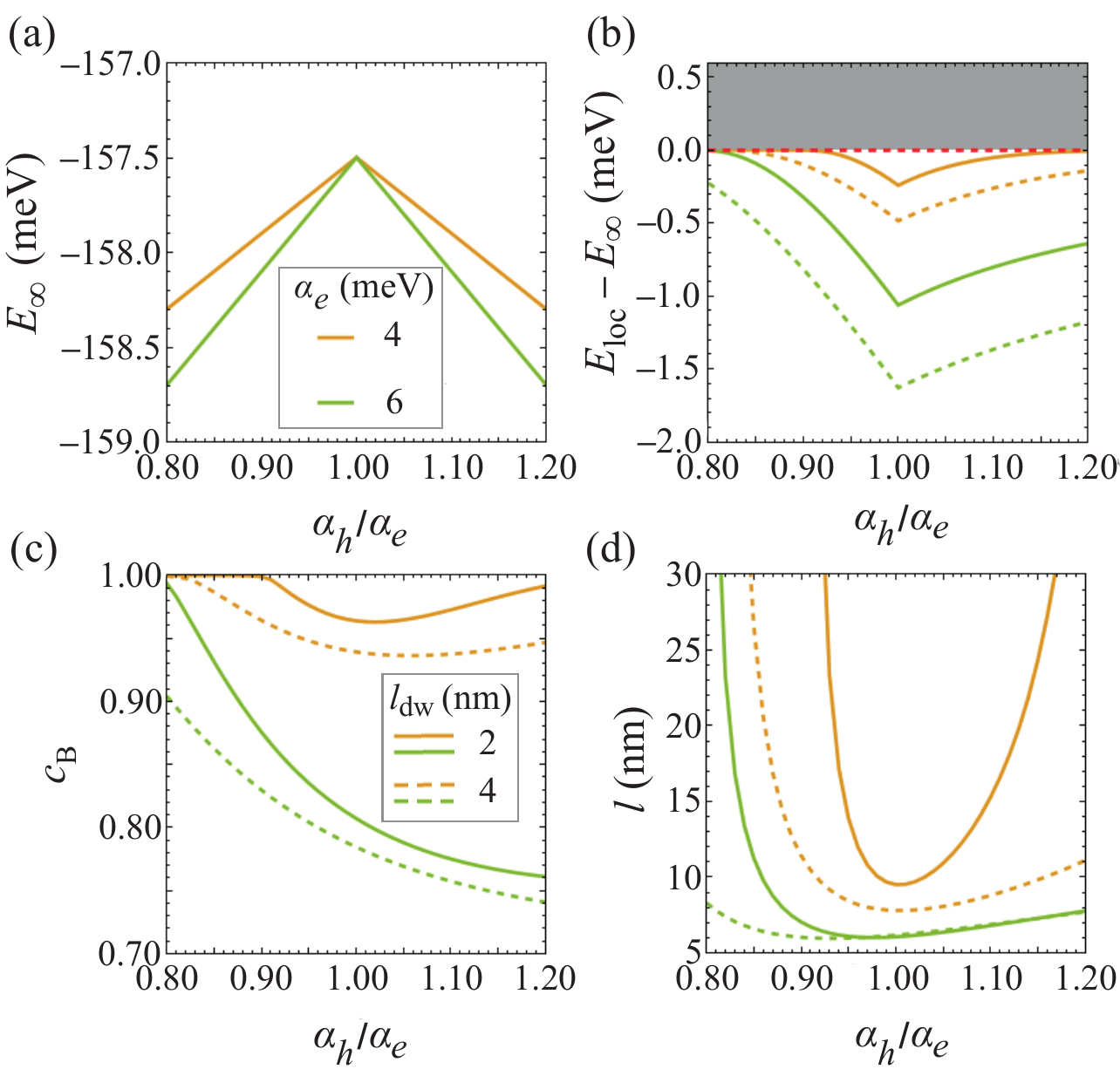}
    \caption{Properties of the localized exciton state at various strengths of electron and hole exchange coupling to magnetization field.
    (a) Energy of the bright exciton state outside of the domain wall region versus the exchange coupling rate $\alpha_h$ for the hole (measured in units of $\alpha_e$). 
    (b) The exciton localization energy $E_{\rm loc}$ relative to the energy of the delocalized state $E_{\infty}$. 
    The grey shaded region indicates continuum of delocalized states.
    (c) Fraction of bright exciton in the localized state. 
    (d) Localization width $l$ of the trapped exciton defined as the root second-moment width of the bright exciton state $|\chi_{\rm B}|^2$ \cite{SM}. 
    For all panels, the green [orange] lines correspond to $\alpha_e=$ 6 [4] meV, solid [dashed] lines correspond to the DW width $l_{\rm dw} = $ 2 [4] nm.
    }
    \label{fig:local-det}
\end{figure}

\begin{figure}
    \centering   
    \includegraphics[width=0.99\linewidth]{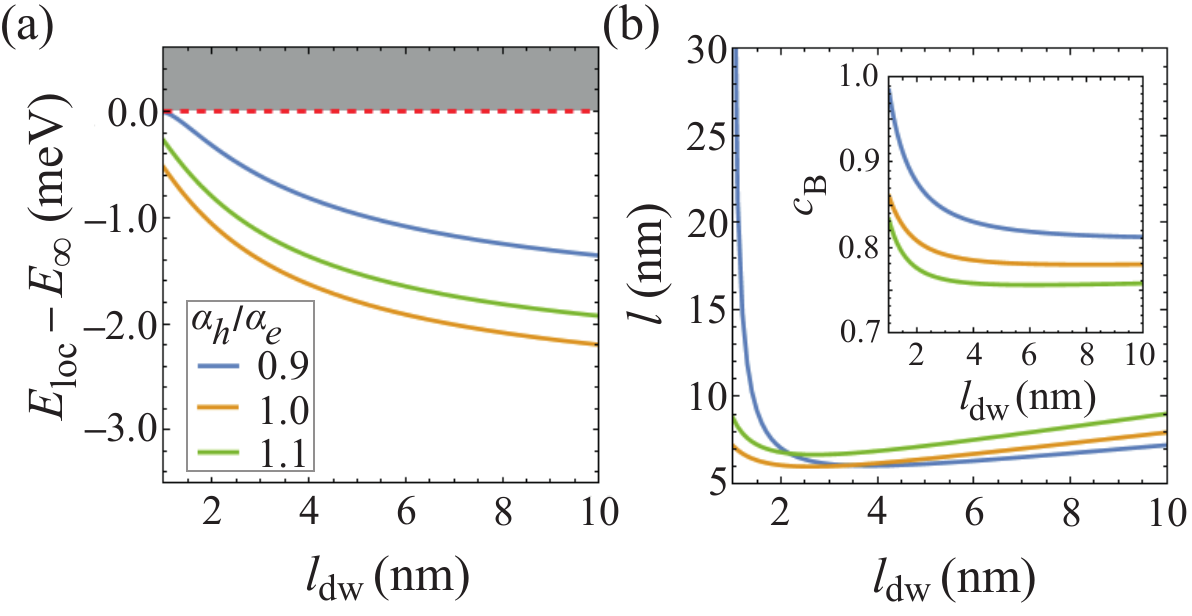}
    \caption{(a) Energy and (b) localization width of the localized exciton vs the domain wall width parameter $l_{\rm dw}$ at $\alpha_e = 6$~meV.  The inset shows the fraction of bright state in the trapped exciton state.
    }
    \label{fig:local-width}
\end{figure}

For the efficient confinement, the trapped state must lie below free exciton energy, which is simply $E_{\infty}=\Tilde{E}_{\rm b}^{\rm B}-\left|\alpha_h-\alpha_e\right|$  far away from the DW, see Fig.~\ref{fig:local-det}(a).
Here $\Tilde{E}_{\rm b}^{\rm B}$ is the binding energy of bright exciton in the absence of magnetization field.
Fig.~\ref{fig:local-det} presents the detailed analysis of exciton localization dependence on the exchange coupling parameters $\alpha_{e,h}$ which can be tuned in a wide range by means of electrical gating and the layer twist angle \cite{Zollner2019}. 
In fact, manipulation of the layer twist angle can substantially modify the exchange interaction strength as it was discussed in Refs.~\cite{Zollner2023,Ren2022}.
Here we account for a realistic case of asymmetric exchange interaction, $\alpha_e \neq \alpha_h$ and use the ratio {$\gamma = \alpha_h/\alpha_e$}, as a scanning parameter.

The localization energy $E_{\rm loc} - E_{\infty}$ is shown in Fig.~\ref{fig:local-det}(b). 
Note that the strong imbalance of the electron and hole exchange strength is detrimental to localization. 
Indeed, the depth $E_{\rm trap}$ of the effective trap \eqref{eq:EEff}, which can be approximated by $E_{\rm trap} \approx \gamma \alpha_e^2/\left(2\Delta\right) + \left(\gamma -1\right)^2\Delta/2 - \alpha_e\left| 1 - \gamma \right| $ \cite{SM}, is maximized at $\gamma=1$ corresponding to the symmetric exchange coupling. Here $\Delta$ is the energy gap between the unperturbed dark and bright excitons.
In this optimal configuration the localization width is about 10~nm, see Fig.~\ref{fig:local-det}(d). 
The increasing $\alpha_h/\alpha_e$ ratio leads to reduction of localization energy together with rapid increase of localization width, with the subsequent delocalization. 

Confined excitons can be detected from their photoluminescence signal. 
The relevant optical response is characterized by the fraction and the radiative recombination rate of the bright state. In turn, the exciton radiative recombination is governed by the modulus of wave function of exciton relative dynamics taken at origin of relative coordinate, i.e. when the positions of the electron and hole overlap \cite{Burstein}.
The spatial variation  of the normal-to-plane magnetization field at the DW core stretches exciton, with the reduction of the electron–hole overlap and respective suppression of exciton radiative recombination rate. 
Indeed, the electron-hole pair experiences an exchange-induced Zeeman shift $U = \alpha_e M_z(\mathbf{r}_e) - \alpha_h M_z(\mathbf{r}_h)$ whose gradient points across DW and tends to localize carriers at the complementary sides of DW. 
The corresponding effective polarizing field $F \propto \partial U/\partial x$ is below few tens of kV$/$cm$^2$. 
Therefore, this effect is minor and can be safely neglected, that is confirmed by direct computations. 
Thus, the luminescence efficiency of the trapped state is determined solely by the bright exciton fraction $c_{\rm B}  = \int|\chi_{\rm B}|^2{\rm d}X$ shown in Fig.~\ref{fig:local-det}(c).

Note that characteristics of the DW internal structure affect localization phenomenon as well.
The increase of the DW width $l_{\rm dw}$ enhances localization efficiency, c.f. the solid and dashed curves in Figs.~\ref{fig:local-det}(b),(d). The width of DW can be manipulated in the magnetic bilayer structures by tuning the stacking angle \cite{Soriano2022}.
The extended $l_{\rm dw}$-dependence is summarized in Fig.~\ref{fig:local-width}.
At the limit of narrow DW, the exciton pinning is quenched, and its localization energy reaches zero, see Fig.~\ref{fig:local-width}(a).
As the DW width increases, the ground state energy gradually saturates while localization width exhibits nonmonotonic behavior shown in Fig.~\ref{fig:local-width}(b). 
This feature can be explained in terms of the effective trapping potential $E_{\rm X}^{\rm eff}$ whose width correlates with $l_{\rm dw}$.
In the presence of narrow DW, the exciton wave function gets delocalized before merging with the continuum of free propagating states. 
In the opposite limit, the exciton state falls on the bottom of the trap, and its size increases as the trap width grows together with the DW thickness $l_{\rm dw}$. In between, the optimal localization regime is achieved. 
At the feasible exchange strength below 10 meV, the best exciton localization conditions are expected on the few nanometer-wide DWs, -- see Fig.~\ref{fig:local-width}(b), which is consistent with theoretical predictions \cite{Abdul-Wahab2020,Abdul-Wahab2021}. 
The corresponding reduction of the exciton oscillator strength is about twenty percent, see the inset to Fig.~\ref{fig:local-width}(b) and Fig.~\ref{fig:local-det}(c).

{\it Conclusion and outlook}. We have demonstrated a new mechanism of the 1D localization of excitons in a TMD monolayer placed in the proximity of the monolayer ferromagnet. 
The localization emerges at the magnetic domain wall due to the in-plane magnetization which mixes spin-bright and spin-dark excitons. 
The proposed mechanism is compatible with various combinations of atom-thick magnets and Mo-based monolayers, i.e. MoSe$_2$, MoS$_2$.
In tungsten based monolayers the large spin-orbit splitting strongly suppresses the state mixing, preventing thus the formation of an effective trap, and ultimately the localized state.

The predicted effect can be employed for the purely optical mapping of the DWs formed in 2D ferromagnets. 
The proposed experimental scheme is based on a non-resonant optical excitation, implying the formation of both free and localized bright excitons. 
The emission from excitons pinned on the DW will be redshifted with respect to free excitons, whereas its intensity reduction due to the state mixing will be expectedly negligible. 
Since the DW-localized exciton emission can be spectrally isolated from the pristine exciton line, one may resort to the photoluminescence imaging techniques allowing deeply subwavelength positioning of the DW. In a recent paper \cite{Abramov2023} this technique was used to define the position of strain-induced single photon emitters in WSe$_2$ with $\sim$5~nm resolution.
Alternatively, the required resolution may be achieved with the use of scanning near-field optical microscopy. 

Observation of the proposed effect is generally hindered by the presence of impurities, structural defects, and other undesired side effects. 
Therefore, the experimental observation requires the use of good quality samples with the minimized number of structural defects, which can be particularly achieved using hBN-encapsulated samples.
Moreover, the emission from the DW-localized excitons can be easily distinguished from the defect-induced single point emitters due its stripe-like shape defined by the structure of underlying magnetic DW.   
The predicted 1D localization phenomenon opens perspectives for routing of the excitons along the DWs which paves the way towards novel application of vdW heterostructures in spintronics.

\section*{Acknowledgments}
This work was supported by the Ministry of Science and Higher Education of Russian Federation, goszadanie no. 2019-1246.
V.S. acknowledges the support of ‘Basis’ Foundation (Project No. 22-1-3-43-1). The work of I.C. (developing formalism, numerical analysis, text writing) was supported by Russian Science Foundation (Project No. 22-72-00061).


\bibliography{ExLocDWbibl}

\end{document}